\documentclass[runningheads]{llncs}
\usepackage[T1]{fontenc}
\usepackage{array}
\usepackage{multirow}
\usepackage{graphicx,verbatim}
\usepackage{tikz}
\begin{document}
\title{MOST: MR reconstruction Optimization for multiple downStream Tasks via continual learning}
%
% \begin{comment}  %emoved for anonymized MICCAI 2025 submission

\author{Hwihun Jeong\inst{1,2} \and
Se Young Chun\inst{1,3} \and
Jongho Lee\inst{1}}
\authorrunning{H. Jeong et al.}
\institute{Department of ECE, Seoul National University, Seoul, Korea \\
\email{\{hwihuni, sychun, jonghoyi\}@snu.ac.kr} \and
Brigham and Women's Hospital, Harvard Medical School, Boston, USA \and
INMC \& IPAI, Seoul National University, Seoul, Korea \\
}

% \end{comment}

% \author{Anonymized Authors}  %% Added for anonymized MICCAI 2025 submission
% \authorrunning{Anonymized Author et al.}
% \institute{Anonymized Affiliations \\
%     \email{email@anonymized.com}}

\maketitle              % typeset the header of the contribution
\begin{abstract}
Deep learning-based Magnetic Resonance (MR) reconstruction methods have focused on generating high-quality images but often overlook the impact on downstream tasks (e.g., segmentation) that utilize the reconstructed images. Cascading separately trained reconstruction network and downstream task network has been shown to introduce performance degradation due to error propagation and domain gaps between training datasets. To mitigate this issue, downstream task-oriented reconstruction optimization has been proposed for a single downstream task. Expanding this optimization to multi-task scenarios is not straightforward. In this work, we extended this optimization to sequentially introduced multiple downstream tasks and demonstrated that a single MR reconstruction network can be optimized for multiple downstream tasks by deploying continual learning (MOST). MOST integrated techniques from replay-based continual learning and image-guided loss to overcome catastrophic forgetting. Comparative experiments demonstrated that MOST outperformed a reconstruction network without finetuning, a reconstruction network with naïve finetuning, and conventional continual learning methods. 
\keywords{Continual learning \and MRI reconstruction \and Downstream task}
% Authors must provide keywords and are not allowed to remove this Keyword section.

\end{abstract}
\section{Introduction}
Magnetic Resonance Imaging (MRI) has gained considerable prominence in clinical practice. The advent of deep learning has led to innovations in MRI such as disease classification \cite{cuingnet2011automatic,li2015robust,peran2018mri,suk2014hierarchical}, tumor or lesion segmentation \cite{chen2017fully,havaei2017brain}, acquisition \cite{shin2021deep}, and image processing \cite{yoon2018quantitative}. In particular, the scan time reduction has been a focus of innovation, and various reconstruction networks have been proposed to produce high-quality images from highly undersampled k-space data \cite{aggarwal2018modl,hammernik2018learning,zbontar2018fastmri}. When evaluating reconstructed images, the gold standard is the assessment by radiologists, which is often impractical due to high cost. Consequently, many studies rely on quantitative metrics such as the Structural Similarity Index (SSIM) and Peak Signal-to-Noise Ratio (PSNR) \cite{muckley2021results,zbontar2018fastmri}. Despite the utility, such metrics have limitations in reflecting the evaluation of radiologists. Moreover, when the reconstruction network is connected to a downstream task network (e.g., segmentation network), optimizing these metrics has been shown to deliver suboptimal outcomes in the downstream task \cite{desai2021international,desai2022skm}. As a result, efforts have been made to design a “downstream task-oriented” reconstruction network, optimizing it based on the performance of downstream tasks \cite{fan2018segmentation,sun2019joint,wang2021one,wu2024learning}. So far, however, these methods are designed for a single downstream task (Fig. \ref{fig:mostconcept}a). Hence, when multiple downstream tasks exist, the current solution is to develop multiple reconstruction networks each optimized for individual tasks, which may not be practical.

In this study, we propose an approach for the MR reconstruction Optimization for multiple downStream Tasks (MOST). Simple expansion from single-task to multi-task optimization is not straightforward \cite{sener2018multi}. Therefore, we consider a realistic scenario where the reconstruction network is sequentially finetuned for pretrained downstream task networks, assuming limited access to the training dataset of previously trained tasks. (Fig. \ref{fig:mostconcept}b) This sequential finetuning strategy reduces the computational burden of processing the entire dataset for all previous tasks and handles real-world environments where multiple downstream tasks are progressively developed and introduced.

When sequentially training a reconstruction network for multiple downstream tasks, one may consider naïve finetuning for the tasks. However, this approach can lead to a well-known issue of catastrophic forgetting, where performance on previously optimized tasks degrades substantially \cite{aljundi2018memory,kirkpatrick2017overcoming,li2019learn,li2017learning,loo2020generalized}. To overcome this issue and sustain the performances for the multiple downstream tasks, we propose to adopt the continual learning \cite{buzzega2020dark,chaudhry2019tiny} and tailor it to our approach. 

\section{Method}
\subsection{Finetuning of downstream task-oriented MR reconstruction network}
Accelerated MRI reconstruction aims to produce high-quality images from undersampled k-space data, making them comparable to fully sampled data. Traditionally, an MR reconstruction model denoted as $f_R$ with parameter $\theta$ is trained using the fidelity loss function $\mathcal{L}_R$ such as SSIM loss:
\begin{equation} \theta^* = \arg \min_\theta \mathcal{L}_R\left(f_R(x;\theta),y\right),  \label{eq:fidelrecon} \end{equation}
where $x$ represents an aliased image, and $y$ represents an aliasing-free label image. 

\begin{figure}[t]
\centering
  \includegraphics[width=1\linewidth]{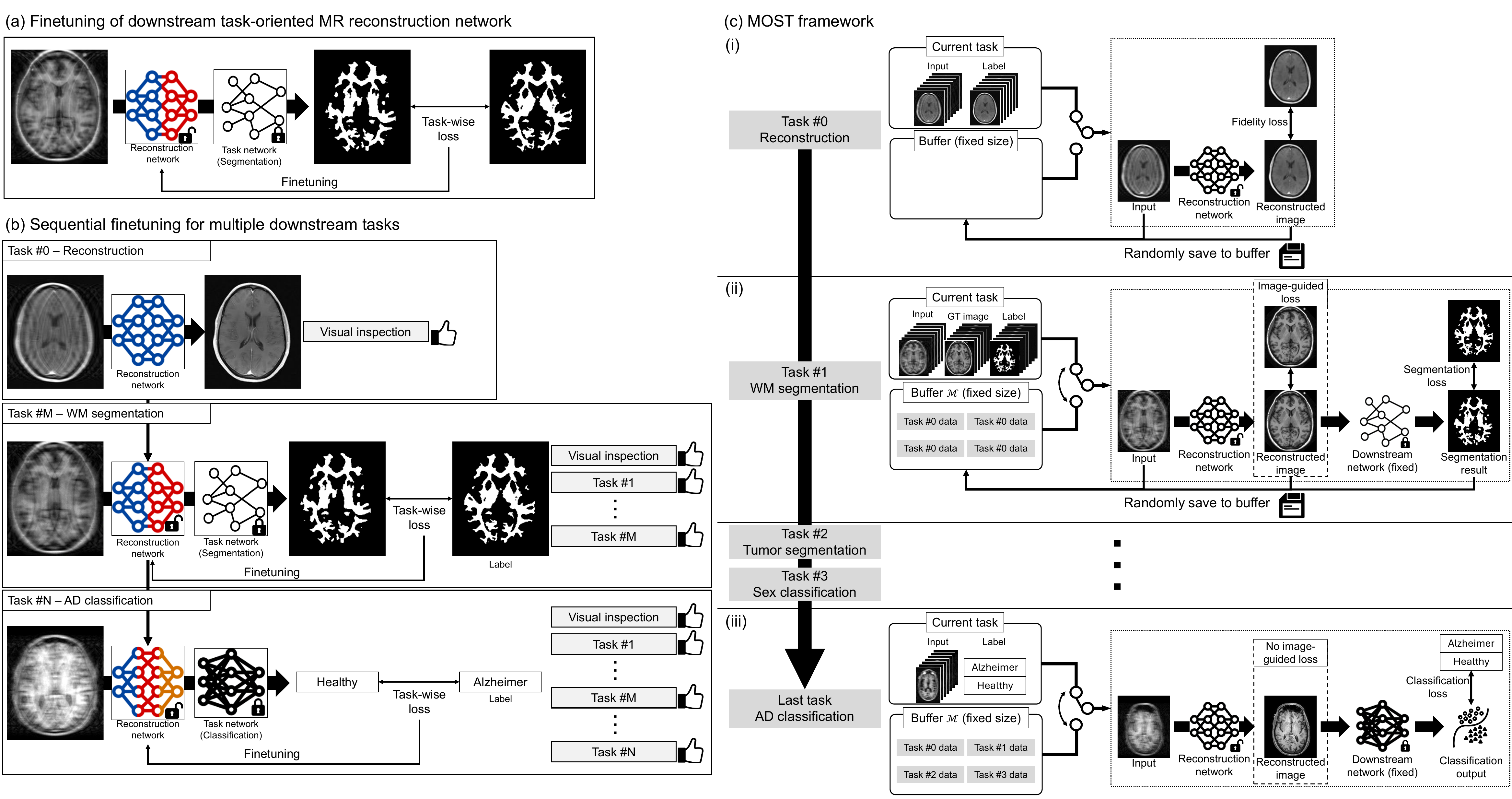}
      % \caption{(a) Combining individually trained reconstruction and downstream networks may result in performance degradation, stemming from error propagation or domain gap issues. Task-oriented finetuning can mitigate these challenges. (b) In scenarios with progressively developed downstream tasks, sequential finetuning of the reconstruction network can be an effective solution. However, naïve finetuning can lead to catastrophic forgetting. (c) MOST addresses the issue of catastrophic forgetting. As finetuning advances, a fixed-size buffer stores input-output pairs, with newly added pairs replacing some of the old ones. Subsequent finetuning stages leverage this buffer data to mitigate catastrophic forgetting. Moreover, the image guided loss is introduced when finetuning for the segmentation task.}
      \caption{(a) Combining individually trained reconstruction and downstream networks may result in performance degradation. Downstream task-oriented finetuning can mitigate these challenges. (b) In scenarios with progressively developed downstream tasks, sequential finetuning of the reconstruction network can be an effective solution. However, naïve finetuning can lead to catastrophic forgetting. (c)  Overview framework of the MOST}
\label{fig:mostconcept}
\end{figure}

However, when this reconstructed network is used for a downstream task such as segmentation and classification, a simple cascade of independently trained reconstruction network and downstream network might not be optimal. For example, the reconstructed images may have different characteristics from the images trained for the downstream network due to the imperfection of the reconstruction network or domain gaps between the two training datasets. To address this issue, finetuning of the reconstruction network with end-to-end data (cf., data pairs of aliased image and downstream task label) using a corresponding loss function is a feasible approach (Fig. \ref{fig:mostconcept}a) \cite{fan2018segmentation,sun2019joint,wang2021one,wu2024learning}:
\begin{equation}
    \theta^*_{task} = \arg \min_\theta \mathcal{L}_d\left(f_d(f_R(x;\theta)),z\right),
    \label{eq:taskrecon} 
\end{equation}
where $f_d$ is the downstream task network, $\mathcal{L}_d$ is the corresponding loss function (e.g., cross-entropy for classification), and $z$ is the label (e.g., classification label).

\subsection{Sequential finetuning for multiple downstream tasks}
When multiple downstream tasks exist, the task-oriented optimization for multiple downstream tasks ($t=1,2,3,\cdots,T$) can be rewritten as follows: 
\begin{equation}
    \theta^*_{multi-task} = \arg \min_\theta \sum_t \mathcal{L}_{dt}\left(f_{dt}(f_R(x_t;\theta)),z_t\right),
    \label{eq:multi} 
\end{equation}
where $\mathcal{L}_{dt}$ is the loss function of $t$-th downstream tasks, and $f_{dt}$ is the downstream task network. The end-to-end dataset of finetuning for $t$-th downstream task $D_t=\{x_t^i,z_t^i \}_{i=1}^{n_t}$ includes an aliased image $x$ and a downstream task label $z$ with the size of $n_t$. However, such multi-task learning is difficult to be optimized \cite{sener2018multi} and does not adapt well to real-world applications. In this work, we propose that single reconstruction network can be sequentially finetuned for multiple downstream tasks (Fig. \ref{fig:mostconcept}b). Our scenario assumes limited access to previously trained datasets due to privacy concerns and computational costs. Initially, the reconstruction network $f_R$ is trained exclusively for image reconstruction using a loss function ($\mathcal{L}_R$) and a reconstruction dataset ($D_R$) as written in Eq. \ref{eq:fidelrecon}. Then, we sequentially finetune the reconstruction network for multiple downstream tasks:
\begin{equation}
    \theta^*_{t} = \arg \min_\theta \mathcal{L}_{dt}\left(f_{dt}(f_R(x_t;\theta)),z_t\right).
    \label{eq:taskconrecon} 
\end{equation}
Finetuning process exclusively targets the reconstruction network while downstream networks are fixed. When finetuning for the $t$-th downstream task, we assume that only a small subset of the previous downstream task dataset ($D_p, p<t$) is available.

\subsection{MOST}
In our proposed approach, MOST, replay-based continual learning is combined with an image-guided (IG) loss to prevent catastrophic forgetting during sequential finetuning for multiple downstream tasks (Fig. \ref{fig:mostconcept}c).
For the replay, a fixed-size buffer ($\mathcal{M}$) is maintained to store a subset of input-output pairs from previous tasks. After finishing each task, we randomly select input-output pairs from that task and add them to the buffer. The buffer size remains constant throughout the finetuning process, with newly added data pairs replacing older ones. Hence, the number of data in the buffer is kept the same across the tasks. 

% \begin{figure}
% \centering
%   \includegraphics[width=0.6\linewidth]{Fig/Fig2.pdf}
%       \caption{Our MOST addresses the issue of catastrophic forgetting during multi-task finetuning. As finetuning advances, a fixed-size buffer stores input-output pairs, with newly added pairs replacing some of the old ones. Subsequent finetuning stages leverage this buffer data to mitigate catastrophic forgetting. Moreover, the image guided loss is introduced when finetuning for the segmentation task.}
% \label{fig:mostmethod}
% \end{figure}

Unlike conventional replay-based continual learning methods \cite{buzzega2020dark,cha2021co2l,chaudhry2019tiny,prabhu2020gdumb}, which generally combine the data of the current and previous tasks in the same mini-batch, our method presents a unique challenge in creating a combined mini-batch because the input data structure and label type can be different among tasks (e.g., segmentation: 2D input and 2D label vs. classification: 3D input and binary label). To address this challenge, data from the past task in the buffer is used in every $K$ iteration during the finetuning process (Fig. \ref{fig:mostconcept}c-ii and \ref{fig:mostconcept}c-iii). The stored data are used in a round-robin fashion across the tasks.

In downstream tasks that generate images (e.g., segmentation tasks), we introduce an image-guided loss, $\mathcal{L}_{IG}$, to further enhance the finetuning process (Fig. \ref{fig:mostconcept}c-ii).  This loss, which is calculated using a reconstruction loss $\mathcal{L}_R$, computes the difference between the intermediate reconstructed image $f_R (x_t;\theta)$ and the aliasing-free image $y_t$ to enhance the performance of the reconstruction network: 

% L_IG=L_R (f_R (x_t;θ),y_t ).                    (5)
\begin{equation}
    \mathcal{L}_{IG} = \mathcal{L}_R(f_R(x_t;\theta),y_t).
    \label{eq:kdig} 
\end{equation}
The reconstructed image is saved to the buffer to compute the image-guided loss for previous tasks. %The overall pipeline of MOST is summarized in Algorithm \ref{al:MOST}. 

% \begin{algorithm}[t]
%    \caption{MOST pipeline}
%     \textbf{Initialize}: datasets $D$, reconstruction model $f_R$, downstream models $f_d$, loss functions $\mathcal{L}$, learning rate $\lambda$
    
%    \begin{algorithmic}[1]
%    \State{$\mathcal{M} \leftarrow\{ \}$ }
%    \For{$D_t$ in $D$}
%    \For{$k$,$x_t$,$y_t$,$z_t$ in enumerate($D_t$)}
%    \If{$exist(y_t)$}
%    \State{$\mathcal{L}_{KDIG} = \mathcal{L}_R(f_R(x_t;\theta),y_t)$}
%    \State{$\theta \leftarrow \theta + \lambda \nabla ( \mathcal{L}_t\left(f_{dt}(f_R(x_t;\theta)),z_t\right) + \mathcal{L}_{KDIG}) $}
%    \Else
%    \State{$\theta \leftarrow \theta + \lambda \nabla ( \mathcal{L}_t\left(f_{dt}(f_R(x_t;\theta)),z_t\right)$}
%    \EndIf
%    \If{$modular(k, K) == 0$}
%    \State{$x', y', z' \leftarrow \mathcal{M}$}
%    \State{$\mathcal{L}_{KDIG} = \mathcal{L}_R(f_R(x';\theta),y')$}
%    \State{$\theta \leftarrow \theta + \lambda \nabla \mathcal{L}_t\left(f_{d}(f_R(x';\theta)),z'\right) +\mathcal{L}_{KDIG})$}
%    \EndIf
%    \EndFor
%    \State{$\mathcal{M} \leftarrow Reservoir\big(\mathcal{M}, Sample(D_t)\big)$}
   
%    \EndFor
%    \end{algorithmic}
%    \label{al:MOST}
% \end{algorithm}

\section{Experiments and Results}
\subsection{Dataset and implementation details}
We used T1-weighted images from multiple datasets, including FastMRI \cite{zbontar2018fastmri}, OASIS 1 \cite{marcus2007open}, BraTS \cite{menze2014multimodal}, IXI \cite{ixi}, and ADNI \cite{jack2008alzheimer} for image reconstruction, white matter (WM) segmentation, tumor segmentation, sex classification, and Alzheimer's disease (AD) classification, respectively. To generate undersampled k-space data, we applied a forward model to the fully sampled image assuming single-coil acquisition at the acceleration factor of 4. For the segmentation and classification datasets, we divided the dataset into two parts: one for the initial downstream task network and the other for downstream task-oriented finetuning.

For the reconstruction network, an single coil version end-to-end variational network \cite{hammernik2018learning,zbontar2018fastmri} is deployed. The U-net architecture \cite{ronneberger2015u} was used for the network part of the model. We also employed the same U-net architecture for the downstream task of segmentation tasks, and we utilized a CNN for classification tasks. The cross-entropy loss served as a loss function for segmentation and classification tasks. The finetuning process for each downstream task was conducted for maximum 5 epochs, and the network with the minimum validation loss was chosen. 
%For the classification tasks, individual slices of a 3D volume were processed through the reconstruction network and then the reconstructed 3D volume was inputted to the classification network. 
We utilized buffer data every three iterations ($K=3$).

We assessed the quality of the reconstructed images using the structural similarity index (SSIM), segmentation results with the Dice similarity coefficient (DICE), and classification performance with the area under the curve (AUC) after completing finetuning for the last downstream task (Last Metric; LM). Additionally, forgetting measures (FM) were computed to quantify how much a model forgot previously learned tasks as it learned new ones. The forgetting measures were calculated by the difference between the best metric and the worst metric of a task obtained during sequential finetuning.

\subsection{Performance evaluation}
We compared the performance of our MOST approach with a reconstruction network without downstream task-oriented finetuning (Without Finetuning) and a reconstruction network with naïve sequential finetuning (Naïve Finetuning). The buffer size was fixed to 10 subjects. The order of the tasks was WM segmentation, tumor segmentation, sex classification, and finally AD classification.  
\begin{figure}[t]
\setlength{\tabcolsep}{0pt}
\centering
\fontsize{6pt}{0pt}\selectfont
\begin{tabular}{>{\centering\arraybackslash}m{0.03\textwidth}>{\centering\arraybackslash}m{0.06\textwidth}>{\centering\arraybackslash}m{0.125\textwidth}>{\centering\arraybackslash}m{0.125\textwidth}>{\centering\arraybackslash}m{0.125\textwidth}>{\centering\arraybackslash}m{0.125\textwidth}>{\centering\arraybackslash}m{0.125\textwidth}}

&&Input & Label & Without Finetuning & Naïve Finetuning & MOST (ours)\\
\multirow{2}{*}{\rotatebox[origin=l]{90}{Reconstruction}} &  
\rotatebox[origin=l]{90}{Slice 1} &
 \includegraphics[width=0.125\textwidth]{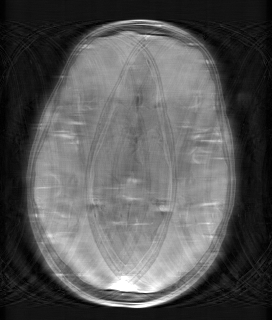}&
 \includegraphics[width=0.125\textwidth]{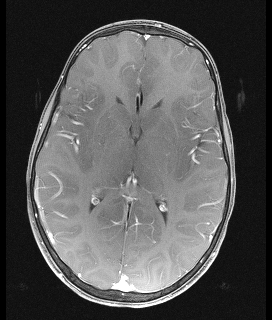}&
 \includegraphics[width=0.125\textwidth]{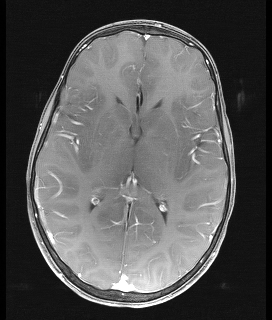}&
 \includegraphics[width=0.125\textwidth]{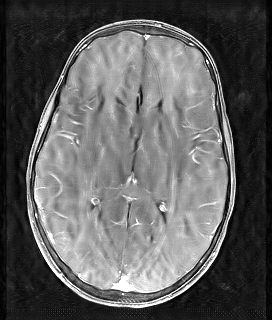}&
 \includegraphics[width=0.125\textwidth]{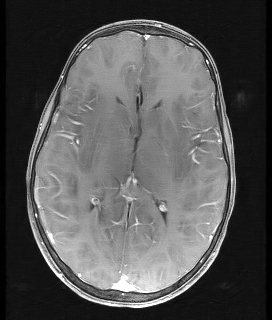}\\
 & \rotatebox[origin=l]{90}{Slice 2} &
 \includegraphics[width=0.125\textwidth]{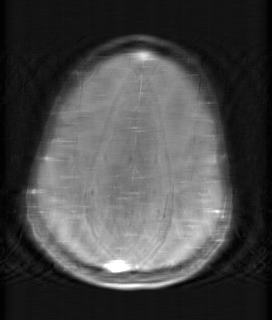}&
 \includegraphics[width=0.125\textwidth]{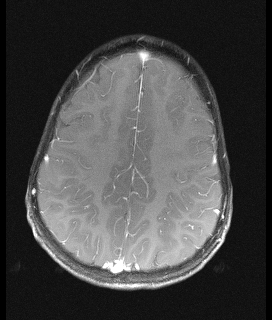}&
 \includegraphics[width=0.125\textwidth]{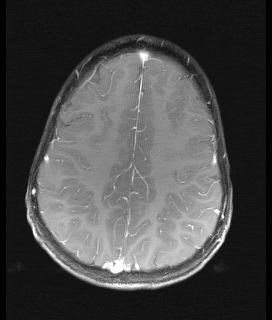}&
 \includegraphics[width=0.125\textwidth]{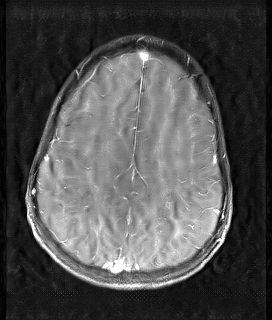}&
 \includegraphics[width=0.125\textwidth]{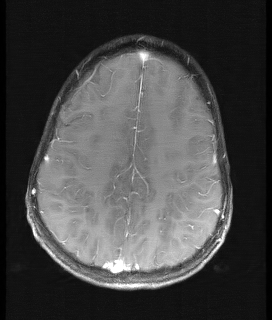}\\
 
 \multirow{2}{*}{\rotatebox[origin=l]{90}{WM segmentation}}  & 
 \rotatebox[origin=l]{90}{Reconstruction} &
 \includegraphics[width=0.125\textwidth]{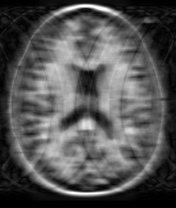}&
 \includegraphics[width=0.125\textwidth]{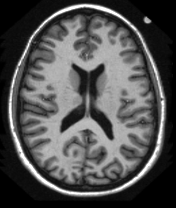}&
 \includegraphics[width=0.125\textwidth]{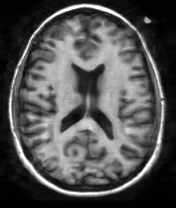}&
 \includegraphics[width=0.125\textwidth]{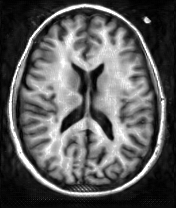}&
 \includegraphics[width=0.125\textwidth]{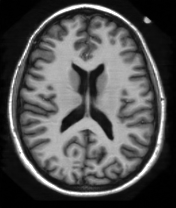}\\
& \rotatebox[origin=l]{90}{Segmentation} & &
 \includegraphics[width=0.125\textwidth]{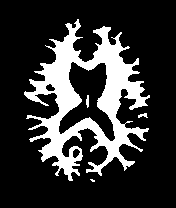}&  
\begin{tikzpicture}
\node[anchor=north east,inner sep=0] (image) at (0,0) {\includegraphics[width=0.125\textwidth]{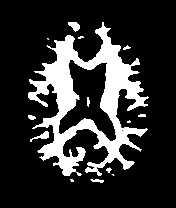}}; % 이미지 불러오기
    \begin{scope}[x={(image.south east)},y={(image.north west)}] % 이미지를 (0,0)~(1,1) 좌표로 맞춤
        \draw[red, very thick, ->] (0.05,0.73) -- (0.15,0.63) ;
    \end{scope}
    \begin{scope}[x={(image.south east)},y={(image.north west)}] % 이미지를 (0,0)~(1,1) 좌표로 맞춤
        \draw[red, very thick, ->] (0.6,0.43) -- (0.7,0.53) ;
    \end{scope}
\end{tikzpicture}&
\begin{tikzpicture}
\node[anchor=north east,inner sep=0] (image) at (0,0) {\includegraphics[width=0.125\textwidth]{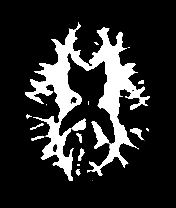}}; % 이미지 불러오기

    \begin{scope}[x={(image.south east)},y={(image.north west)}] % 이미지를 (0,0)~(1,1) 좌표로 맞춤
        \draw[red, very thick, ->] (0.35,0.43) -- (0.45,0.53) ;
    \end{scope}
    \begin{scope}[x={(image.south east)},y={(image.north west)}] % 이미지를 (0,0)~(1,1) 좌표로 맞춤
        \draw[red, very thick, ->] (0.6,0.4) -- (0.7,0.5) ;
    \end{scope}
\end{tikzpicture}&
 \includegraphics[width=0.125\textwidth]{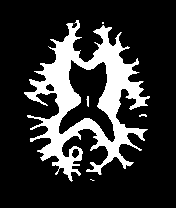}\\

     \end{tabular}
     \caption{The results of the reconstruction and WM segmentation before and after the downstream task-oriented finetuning are illustrated. }
   \label{fig:ill}
 \end{figure}
 
 \begin{table}[t]
   \caption{The last metric (LM) and forgetting measure (FM) of SSIM for the reconstruction, DICE for the segmentation, and AUC for the classification are reported.}
   \label{table:performance}
 \centering
  \setlength{\tabcolsep}{1pt}
\fontsize{8pt}{9pt}\selectfont
\begin{tabular}{c|cccccccccc}
   \hline          
   & \multicolumn{2}{c}{Reconstruction}& \multicolumn{2}{c}{WM seg} &\multicolumn{2}{c}{Tumor seg} & \multicolumn{2}{c}{Sex class} &  \multicolumn{2}{c}{AD class} \\
              
            & \multicolumn{2}{c}{SSIM} & \multicolumn{2}{c}{DICE} &  \multicolumn{2}{c}{DICE} & \multicolumn{2}{c}{AUC} & \multicolumn{2}{c}{AUC}  \\
              \hline 
              &  LM($\uparrow$) & FM($\downarrow$)& LM($\uparrow$) & FM($\downarrow$)& LM($\uparrow$) & FM($\downarrow$)& LM($\uparrow$) & FM($\downarrow$)& LM($\uparrow$)& FM($\downarrow$) \\ 
              \hline
   % Recon Label Input   &-    &-    & 0.976  &-    & 0.685     &- & 0.983     &- & 0.810  &- \\
   Without Finetuning  &0.977  &-  & 0.861   &-   & 0.564    &-  & 0.983    &-  & 0.800  &- \\
   Naïve Finetuning    &0.866 & 0.110  & 0.835  & 0.098   & 0.445 & 0.179    & 0.976   & 0.007  & 0.812  & -  \\
   MOST (ours)         &0.971  & \textbf{0.006}   & \textbf{0.938}  & \textbf{0.002}   & \textbf{0.642}  & \textbf{0.003}   & 0.983  &\textbf{ 0.000}   & \textbf{0.813} & - \\
   \hline

   \end{tabular}
   
\end{table}

The qualitative results of the reconstruction and WM segmentation tasks are illustrated in Fig. \ref{fig:ill}. Without downstream task-oriented finetuning (Fig. \ref{fig:ill}; third column), the reconstructed images demonstrate good quality for the reconstruction dataset, but exhibit inferior results for the WM segmentation dataset due to domain gaps or error propagation. Naïve Finetuning results in poor quality in both reconstruction and segmentation, indicating catastrophic forgetting (Fig. \ref{fig:ill}; fourth column). The MOST approach effectively mitigates this issue, producing high-quality results in both tasks (Fig. \ref{fig:ill}; last column).

Table \ref{table:performance} provides the quantitative metrics. Compared to Without Finetuning, MOST successfully improved performance in all four downstream tasks with a minor decrease in reconstruction quality, demonstrating the effectiveness of finetuning. Compared to Naïve Finetuning, MOST exhibited superior LM and FM, demonstrating its effectiveness in preventing catastrophic forgetting.
% Moreover, it can be observed that the metric of MOST is comparable to the Recon Label Input case, which presumably produces the best outcome (see the second and fifth rows in Table \ref{table:performance}).

\begin{table}[t]
   \caption{Result of last metric (LM) and forgetting measures (FM) with three different downstream task orders (a) and comparison result to the other continual learning methods (b) are illustrated.}
   \label{table:ordercomp}
 \centering
  \setlength{\tabcolsep}{1pt}
\fontsize{8pt}{9pt}\selectfont
   \begin{tabular}{c|cccccccccc}
   \hline
          (a)  & \multicolumn{2}{c}{Reconstruction}& \multicolumn{2}{c}{WM seg} &\multicolumn{2}{c}{Tumor seg} & \multicolumn{2}{c}{Sex class} &  \multicolumn{2}{c}{AD class} \\
              
            & \multicolumn{2}{c}{SSIM} & \multicolumn{2}{c}{DICE} &  \multicolumn{2}{c}{DICE} & \multicolumn{2}{c}{AUC} & \multicolumn{2}{c}{AUC}  \\
              \hline 
              &  LM($\uparrow$) & FM($\downarrow$)& LM($\uparrow$) & FM($\downarrow$)& LM($\uparrow$) & FM($\downarrow$)& LM($\uparrow$) & FM($\downarrow$)& LM($\uparrow$)& FM($\downarrow$) \\ 
              \hline 
   Without Finetuning     & 0.977 & - & 0.861 & - & 0.564 & - & 0.983 & - & 0.800 &-      \\
              \hline
               & \multicolumn{10}{c}{Order 1: Recon $\rightarrow$ WM seg $\rightarrow$ Tumor seg $\rightarrow$ Sex class$ \rightarrow$ AD class}\\
              \hline 
   Naïve Finetuning     & 0.866 & 0.110 & 0.835 & 0.098 & 0.445 & 0.179 & 0.976 & 0.007 & 0.812 & -      \\
   MOST (ours)          & 0.971 & \textbf{0.006} &\textbf{ 0.938} & \textbf{0.002} & \textbf{0.642} & \textbf{0.003 } & \textbf{0.983} & \textbf{0.000} & \textbf{0.813} & -       \\

    \hline 
               & \multicolumn{10}{c}{Order 2: Recon $\rightarrow$  Tumor seg $\rightarrow$ AD class $\rightarrow$   WM seg $\rightarrow$ Sex class}\\
              \hline 
   Naïve Finetuning      & 0.843 & 0.133 & 0.930 & 0.002 & 0.453 & 0.181 & 0.977 & - & 0.761 & 0.051       \\
   MOST (ours)          & 0.970 & \textbf{0.007} &\textbf{0.936} &\textbf{ 0.001} & \textbf{0.641} & \textbf{0.010}  & \textbf{0.983} & - & \textbf{0.804} & \textbf{0.012}          \\

    \hline 
               & \multicolumn{10}{c}{Order 3: Recon $\rightarrow$ Sex class $\rightarrow$ AD class $\rightarrow$  Tumor seg  $\rightarrow$ WM seg}\\
              \hline 
   Naïve Finetuning      & 0.791 & 0.186 & \textbf{0.934} &-& 0.421 & 0.207  & 0.956 & 0.031 & 0.761 & 0.053       \\
   MOST (ours)          & 0.959 & \textbf{0.018} &0.920 & - & \textbf{0.598} & \textbf{0.037}  & \textbf{0.983} & \textbf{0.000} & \textbf{0.802} & \textbf{0.007}       \\

   \hline

   \end{tabular}

      \begin{tabular}{c|ccccccccccc}
   \hline
          (b)   & \multicolumn{2}{c}{Reconstruction}& \multicolumn{2}{c}{WM seg} &\multicolumn{2}{c}{Tumor seg} & \multicolumn{2}{c}{Sex class} &  \multicolumn{2}{c}{AD class} \\
              
          & \multicolumn{2}{c}{SSIM} & \multicolumn{2}{c}{DICE} &  \multicolumn{2}{c}{DICE} & \multicolumn{2}{c}{AUC} & \multicolumn{2}{c}{AUC}  \\
              \hline 
              &  LM($\uparrow$) & FM($\downarrow$)& LM($\uparrow$) & FM($\downarrow$)& LM($\uparrow$) & FM($\downarrow$)& LM($\uparrow$) & FM($\downarrow$)& LM($\uparrow$)& FM($\downarrow$) \\ 
              \hline 

   Naïve Finetuning      & 0.866 & 0.110 & 0.835 & 0.098 & 0.445 & 0.179 & 0.976 & 0.007 & 0.812 & -      \\
   EWC \cite{kirkpatrick2017overcoming}    & 0.910 & 0.066 & 0.884 & 0.048 & 0.536 & 0.085 & 0.976 & 0.007 & 0.808 & -    \\
   LWF \cite{li2017learning}   & 0.937 & 0.040 & 0.892 & 0.026 & 0.488 & 0.057 & 0.893 & 0.041 & 0.785  & -      \\
   ER \cite{chaudhry2019tiny}   & 0.953 & 0.024 & 0.859 & 0.073 & 0.479 & 0.154 & 0.965 & 0.018 & 0.810  & -      \\
   DER \cite{buzzega2020dark} & 0.971 & 0.006 & 0.929 & 0.004 & 0.622 & \textbf{0.000} & 0.981 & 0.002 & 0.809  & -     \\
   MOST (ours)         & 0.971 & 0.006 &\textbf{ 0.938} & \textbf{0.002} & \textbf{0.642} & 0.003  & \textbf{0.983} & \textbf{0.000 }& \textbf{0.813} & -       \\
   \hline

   \end{tabular}

\end{table}
\begin{table}[t]
   \caption{(a) MOST with various buffer sizes (the number of subjects) is tested. (b) Ablation study without replay-based continual learning and/or image-guided (IG) loss is illustrated.}
   \label{table:sizeabl}
\centering%\resizebox{\linewidth}{!}
\fontsize{8pt}{9pt}\selectfont
\begin{tabular}{c|cccccccccc}
   \hline
          (a)  & \multicolumn{2}{c}{Reconstruction}& \multicolumn{2}{c}{WM seg} &\multicolumn{2}{c}{Tumor seg} & \multicolumn{2}{c}{Sex class} &  \multicolumn{2}{c}{AD class} \\
              
           buffer size & \multicolumn{2}{c}{SSIM} & \multicolumn{2}{c}{DICE} &  \multicolumn{2}{c}{DICE} & \multicolumn{2}{c}{AUC} & \multicolumn{2}{c}{AUC}  \\
              \hline 
              &  LM($\uparrow$) & FM($\downarrow$)& LM($\uparrow$) & FM($\downarrow$)& LM($\uparrow$) & FM($\downarrow$)& LM($\uparrow$) & FM($\downarrow$)& LM($\uparrow$)& FM($\downarrow$) \\ 
              \hline 
   4   &0.970 &0.007 & 0.933& 0.007 & 0.640 & 0.006 & 0.982& 0.001  & 0.810  & -  \\
   10  &0.971&0.006 & 0.938 & 0.002& 0.642   & 0.003& 0.983& 0.000  & 0.813  & -\\
   50  &0.970&0.007 & 0.937 & 0.003& 0.640 & 0.004 & 0.981 & 0.003  & 0.810 & - \\
   \hline
   \end{tabular}
   \begin{tabular}{cc|cccccccccc}
   \hline
          \multicolumn{2}{c}{(b)}  & \multicolumn{2}{c}{Reconstruction}& \multicolumn{2}{c}{WM seg} &\multicolumn{2}{c}{Tumor seg} & \multicolumn{2}{c}{Sex class} &  \multicolumn{2}{c}{AD class} \\
              
            Replay & IG loss & \multicolumn{2}{c}{SSIM} & \multicolumn{2}{c}{DICE} &  \multicolumn{2}{c}{DICE} & \multicolumn{2}{c}{AUC} & \multicolumn{2}{c}{AUC}  \\
              \hline 
             & &  LM($\uparrow$) & FM($\downarrow$)& LM($\uparrow$) & FM($\downarrow$)& LM($\uparrow$) & FM($\downarrow$)& LM($\uparrow$) & FM($\downarrow$)& LM($\uparrow$)& FM($\downarrow$) \\ 
              \hline 
   X & O  &0.950 &0.026 & 0.883& 0.056 & 0.604& 0.046  & 0.976 & 0.007& \textbf{0.817}  & -\\
   O & X  &0.971 &0.006& 0.929  & 0.004& 0.622  & \textbf{0.000} & 0.981 & 0.002  & 0.809 & - \\
   O & O  &0.971&0.006 & \textbf{0.938} & \textbf{0.002} & \textbf{0.642} & 0.003 & \textbf{0.983}  & \textbf{0.000} & 0.813& -   \\
   \hline
   \end{tabular}
\end{table}

% \subsubsection{Task order test}
 \subsection{Additional evaluation}
\textbf{Task order experiments} were conducted with three different task orders (Table \ref{table:ordercomp}a). Compared to Without Finetuning and Naïve Finetuning, MOST exhibited superior LM and FM in most of the tasks and task orders, showing minimal dependency on the task order. When comparing across the orders, starting with segmentation proved slightly beneficial in improving the performance, as indicated by a lower LM and FM in Order 3 when compared to Orders 1 and 2. 
% \subsubsection{Comparison with other continual learning methods}

We also \textbf{compared the results to conventional continual learning methods} including Elastic Weight Consolidation (EWC) \cite{kirkpatrick2017overcoming}, Learning Without Forgetting (LWF) \cite{li2017learning}, a memory-based method (ER) \cite{chaudhry2019tiny}, and another replay-based method (DER) \cite{buzzega2020dark}. The results (Table \ref{table:ordercomp}b) clearly demonstrated that the MOST approach achieved overall superior performance compared to the other continual learning methods. In contrast, EWC, LWF, ER, and DER showed varying degrees of forgetting and performance degradation. These findings highlight the advantages of MOST over the conventional continual learning methods, mostly due to the fact that the image-guided loss can provides more comprehensive information. 

When three \textbf{different buffer sizes} (4, 10, and 50 subjects) were evaluated, the results showed consistent performances as shown in Table \ref{table:sizeabl}a, indicating that MOST performed well even with the buffer size of 4. The unit for the buffer size was in the number of subjects. 

% \subsubsection{Ablation study}
To assess the independent contributions of replay-based continual learning and image-guided loss to the performance of MOST, an \textbf{ablation study} was conducted. Three networks were compared: original MOST, MOST without replay-based continual learning, and MOST without image-guided loss. The ablation study results (Table \ref{table:sizeabl}b) demonstrated that original MOST outperformed MOST without either replay-based continual learning or image-guided loss.

\section{Discussion}
In this work, we demonstrated that a single reconstruction network can be sequentially finetuned for multiple downstream tasks. This method successfully generalized recently proposed single downstream task-oriented reconstruction optimization to multiple downstream tasks. In particular, the application of replay-based continual learning along with the image-guided loss successfully prevented catastrophic forgetting, which was observed in naïve finetuning. 

MOST demonstrated improved performance in all task orders when compared to that of Naïve Finetuning, but it showed little dependency on task order (Table \ref{table:ordercomp}a). In particular, MOST underperformed in Order 3, which may be attributed to the introduction of classification tasks earlier than segmentation tasks. Classification labels contained less detailed information compared to segmentation labels, leading to less effective finetuning.

Our proposed framework is based on the single-coil setting, where the reconstruction model takes single-coil k-space data as the input and produces an alias-free image. The reason for this setting is the limited available end-to-end datasets that include multi-coil k-space data and downstream task labels. 

Current continual learning methods mostly target high-level computer vision tasks like classification or segmentation \cite{hadsell2020embracing}, because catastrophic forgetting is rare in regression tasks. In our scenario, although we employ sequential training for the regression network, the cascaded network ultimately performs classifications or segmentations. Hence catastrophic forgetting occurs in the downstream tasks. If the reconstruction network is merely finetuned with the original reconstruction task for different domains (e.g., images with different hardware or parameter), the effect of catastrophic forgetting may decrease, and continual learning may not be required.

In our approach, we manually trained the downstream task networks instead of adopting off-the-shelf state-of-the-art (SOTA) models, resulting in inferior metrics. This approach was necessary because the SOTA model is trained on the entire training dataset, whereas we required the dataset to be split for downstream network training and finetuning. Moreover, the BraTS dataset contained multimodal MRI data, but we only used the T1-weighted images for consistency across downstream tasks. Additionally, most SOTA models included preprocessing steps on the input image, (e.g., skull stripping), which is challenging to be integrated into MOST.

\section{Conclusion}
We introduced that a single reconstruction network can be sequentially finetuned for multiple downstream tasks and demonstrate that replay-based continual learning and image-guided loss can prevent catastrophic forgetting. 
%Through this approach, we paved the way for the development of a more robust and versatile MR reconstruction network in a realistic clinical setup.

{
    % \small
    
\bibliographystyle{splncs04}
    % \bibliographystyle{ieeenat_fullname}
    \bibliography{egbib}
}

\end{document}